# Nano-imaging of intersubband transitions in van der Waals quantum wells

Peter Schmidt[1], Fabien Vialla[1,2], Simone Latini[3,4], Mathieu Massicotte[1], Klaas-Jan Tielrooij[1], Stefan Mastel[5], Gabriele Navickaite[1], Mark Danovich[6], David A. Ruiz-Tijerina[6], Celal Yelgel[6], Vladimir Fal'ko[6], Kristian Thygesen[3], Rainer Hillenbrand[5,7], Frank H. L. Koppens[1,8]*

[1]ICFO - Institut de Ciències Fotòniques, The Barcelona Institute of Science and Technology, 08860 Castelldefels (Barcelona), Spain
[2]Institut Lumière Matière UMR5306, Université Claude Bernard Lyon1 – CNRS, 69622 Villeurbanne Cedex, France
[3]Center for Atomic-scale Materials Design, Technical University of Denmark, DK-2800 Kongens Lyngby, Denmark
[4]Max Planck Institute for the Structure and Dynamics of Matter, 22761 Hamburg, Germany
[5]CIC nanoGUNE Consolider, 20018 Donostia-San Sebastián, Spain
[6]National Graphene Institute, University of Manchester, Manchester M13 9PL, United Kingdom
[7]IKERBASQUE, Basque Foundation for Science, 48011 Bilbao, Spain
[8]ICREA-Institució Catalana de Recerca i Estudis Avançats, 08010 Barcelona, Spain
*Correspondence to: frank.koppens@icfo.eu

**The science and applications of electronics and optoelectronics have been driven for decades by progress in growth of semiconducting heterostructures. Many applications in the infrared and terahertz frequency range exploit transitions between quantized states in semiconductor quantum wells (intersubband transitions). However, current quantum well devices are limited in functionality and versatility by diffusive interfaces and the requirement of lattice-matched growth conditions. Here, we introduce the concept of intersubband transitions in van der Waals quantum wells and report their first experimental observation. Van der Waals quantum wells are naturally formed by two-dimensional (2D) materials and hold unexplored potential to overcome the aforementioned limitations: They form atomically sharp interfaces and can easily be combined into heterostructures without lattice-matching restrictions. We employ near-field local probing to spectrally resolve and electrostatically control the intersubband absorption with unprecedented nanometer-scale spatial resolution. This work enables exploiting intersubband transitions with unmatched design freedom and individual electronic and optical control suitable for photodetectors, LEDs and lasers.**

Nanoscale confinement of charge carriers gives rise to a rich variety of physical phenomena that are not present in bulk materials, and lies at the heart of many modern optoelectronic applications [1-4]. One striking feature is that confinement leads to quantized energy levels, whose energy spacing increases with spatial confinement. Charge carriers can be confined in three dimensions (0D quantum dots [5]), in two dimensions (1D quantum wires [6]) or in one dimension (2D quantum wells). The latter is arguably the most well-known example and has led to the development of quantum well infrared photodetectors [7] and quantum cascade lasers (QCLs) [8] for mid infrared (MIR) to THz wavelengths. These devices rely on transitions of electrons (or holes) between the quantized states – intersubband transitions [9-11] – of a semiconductor quantum well (QW). In contrast to interband transitions, these intraband transitions exhibit very large oscillator strengths close to unity [9]. Currently, state-of-the-art quantum wells are typically grown by molecular beam epitaxy (MBE) of different III-V semiconductor alloys. This technology suffers from fundamental material limitations: Strict lattice matching conditions limit the available material combinations and the thermal growth causes atomic diffusion and increases interface roughness, restricting the performances of actual devices [12].

Recently, 2D materials and its heterostructures have emerged as a promising platform for electronic and optoelectronic applications [13-17], with particular interest for the 2D semiconductors such as metal or transition metal dichalcogenides (TMDs). Studies on hBN-encapsulated TMDs have shown appealing electrical properties such as mobilities > 30,000 cm$^2$/Vs in MoS$_2$ [18] and the demonstration of the quantum hall effect in WSe$_2$ [19] and InSe [20]. Unique optical features have been reported, including spin-valley locking [21], the valley Hall effect [22], near-unity excitonic reflection [23, 24] and photoluminescence close to the homogenous linewidth [25, 26]. All of these optical phenomena in monolayer 2D materials arise from interband electronic transitions.

However, few-layer 2D materials also form natural QWs – van der Waals QWs – where charge carriers are confined to the nanoscale in one dimension and it is therefore natural to expect intersubband transitions to occur. Cross-section TEM images of van der Waals heterostructures have demonstrated defect free and atomically sharp interfaces [14, 18, 27], enabling the formation of ideal QWs, free of diffusive inhomogeneities. Furthermore, van der Waals QWs do not require epitaxial growth on a matching substrate and can therefore be easily isolated and coupled to other electronic systems such as Si CMOS or optical systems such as cavities and waveguides. Finally, by assembling van der Waals QWs into heterostructures, combined with metals or semimetals such as graphene, it is possible to electrically contact each QW individually [28], allowing for in-situ tuning of resonances and interactions as well as facile carrier injection.

Here, we introduce the concept of intersubband transitions in van der Waals quantum wells, report their first experimental observation, and provide detailed theoretical calculations. We probe the transitions between the quantized states within both the valence and the conduction band by investigating the MIR optical response of doped TMDs. The TMDs were exfoliated on a Si/HfO$_2$ substrate - specifically chosen to avoid contributions from substrate phonons - and electron or hole doping was induced by applying a gate voltage between Si and TMD. Each investigated flake comprised several terraces of different layer number *N*, where charge carriers are quantum confined in the out-of-plane direction within the TMD flake, as illustrated in Figure 1a. As the QWs are not buried inside any material, we can perform scattering scanning near-field optical microscopy (s-SNOM) [29] as an innovative measurement approach to spectrally resolve intersubband absorption resonances with a spatial resolution below 20 nm. We spectrally resolve intersubband resonances for the different QW thicknesses within a single TMD flake by varying the illumination photon energy, as schematically depicted in Figure 1b. Furthermore, we demonstrate in-situ control of the intersubband absorption strength by electrostatically tuning the charge carrier density of the TMD flake. Finally, to show the versatility of intersubband transitions in 2D materials, we demonstrate intersubband absorption in both valence and conduction bands within a single device.

First, we lay the theoretical framework for intersubband transitions in van der Waals QWs. Figure 1c shows the band structure of a 2H-WSe$_2$ crystal with layer number *N* = 5, obtained by using ab-initio density functional theory (DFT) calculations including spin-orbit (SO) coupling. We confirm the validity of our calculations by using two different, independent approaches (see SI chapter I for more details). The calculations reveal the splitting of the lowest conduction (highest valence) band into several subbands. In contrast to monolayer TMD crystals, where both band edges appear at the K-points of the Brillouin zone [30], few layer WSe$_2$ has the valence band edge at the Γ point, and the conduction band edge at the Λ points, which are located close to the middle of the Γ-K segments in the Brillouin zone [31, 32]. This change in the characteristics of the band structure is determined by a strong overlap between the Γ and Λ points wave functions in neighboring monolayers which also determines the z-axis (out-of-plane) effective mass and the $k_z$ dispersion in the bulk layered crystal [32]. For the edge of the valence band, which appears at the 3D Γ point in 2H-WSe$_2$, we can approximate the dispersion of holes as $E \approx \frac{\hbar^2 k_z^2}{2m_z} + \frac{\hbar^2 k_{xy}^2}{2m_{xy}}(1 + \zeta k_z^2)$, where $m_z \approx 1.08\, m_e$ and $m_{xy} \approx 0.7\, m_e$ are the out-of-plane and in-plane effective masses obtained by fitting the bulk dispersion, $\hbar$ is the reduced Planck constant, $k_{xy}$ is the in-plane wave vector and $\zeta \approx -5.45\, \text{Å}^2$ is an anisotropic non-parabolicity factor (see SI chapter I.2 for details).

The splitting of the bands corresponds to the quantized states in the quantum well. The wave numbers for these standing waves in the z-direction are determined by the boundary conditions ($[\pm \nu d \partial_z \psi + \psi]_{z=\pm Nd/2} = 0$, with the monolayer thickness $d$ and a phenomenological parameter $\nu$, see SI chapter I.2 for more details) for the hole wave function $\psi(z)$ at the top and bottom interfaces ($z = \pm Nd/2$). Figures 1d and 1e show these (Λ and Γ point) out-of-plane wave functions for electrons and holes, respectively, which represent a periodic modulation with the same periodicity as the atomic potential multiplied by an envelope function. The boundary conditions gives us $k_z = \pi j/(N + 2\nu)d$, with the subband index $j$. The energy of the $j = 1 \rightarrow j = 2$ intersubband transition in a lightly p-doped film is then given by modified infinite square well transition energies

$$E_{sub} = E_{j=2} - E_{j=1} \approx \frac{1}{(N+2\nu)^2} \frac{3\hbar^2 \pi^2}{2m_z d^2}. \qquad (1)$$

This fits very well with the transition energies obtained from the DFT results, shown in Figures 1f and 1g for electrons and holes, respectively.

Now we turn to the experimental observation of intersubband transitions in van der Waals quantum wells. We specifically chose exfoliated TMD flakes consisting of terraces (see Figure 2), in order to compare the optical response of quantum wells of different thicknesses within the same flake. Observing intersubband transitions in these flakes is highly challenging using conventional far-field illumination [9, 33] since the lateral size of each terrace with a constant thickness is smaller than the infrared diffraction limit, and intersubband transitions only couple to light with an out-of-plane polarization component. To overcome these challenges, we used the s-SNOM and exploited the fact that exfoliated 2D materials naturally form a bare QW. The s-SNOM consists of a metalized atomic force microscope (AFM) tip that is illuminated by a continuous-wave (cw) laser source. The nanoscale concentrated infrared field (hot spot) that is generated around the tip apex interacts in the near-field with the TMD flake underneath the tip. Importantly, the hot spot – of about 20 nm in diameter – contains out-of-plane polarization components and can therefore excite intersubband transitions inside the TMD flake. We collect the light scattered back by the tip, which carries (quantitative) information about the sample's complex permittivity and thus its absorption at the illumination photon energy $E_{ph}$. This allows us to spatially resolve intersubband transitions with an unprecedented resolution of 20 nm and a sensitivity high enough to detect the intersubband transition of a single electron (on average) under the AFM tip. By combining interferometric and tip-modulated detection (Figure 2a), we record a complex scattered optical signal $s_3$ at the third harmonic of the AFM tip oscillation frequency $f_{tip}$, whose absolute value $|s_3|$ increases monotonically with $N$, and whose phase $\varphi_3$ is – in the case of weak resonances – proportional to the optical absorption of the TMD flake [35 - 38] (see SI chapter III). We use $|s_3|$ to identify $N$ across the probed flake as shown in Figure 2b.

In order to isolate the absorption signal of intersubband transitions from background signal and noise sources like laser and interferometer drifts, we modulate the charge carrier density of WSe$_2$ by applying a square wave backgate voltage $V_{BG}$ between the Si wafer and two Cr/Au contacts evaporated on the TMD flake (Figure 2c, see SI chapter IV). Applying $V_{BG}$ modulates the doping of the WSe$_2$ flake between charge neutrality and p-doping, hence populating the first subband in the valence band and allowing hole intersubband transitions. When scanning the AFM tip over a TMD flake with different $N$, we simultaneously monitor $V_{BG}$ and the gate-induced change in the third harmonic phase signal - $\Delta\varphi_3$ - extracting the change of absorption of the WSe$_2$ flake induced by a change in carrier density, as shown in Figure 2d. We note that this gate modulation technique improves the signal/noise ratio, but the generally observed trends presented in this work are also visible without modulation technique.

A spatial map of $\Delta\varphi_3$ for $E_{ph}$ = 117 meV over an area of 13x10 μm$^2$ with a spatial resolution of ~20 nm is shown in Figure 2e. A clearly enhanced absorption is observed in the area of the flake where $N$ = 1 and $N$ = 5. When repeating the same scan at $E_{ph}$ = 165 meV, we only observe absorption in the area where $N$ = 4 (Figure 2f). This distinct behavior originates from two different absorption contributions. i) We attribute the enhanced absorption of the $N$ = 4 and $N$ = 5 regions of the flake to intersubband transitions. Comparing the observed absorption energies with the ab-initio calculations from Figure 1g, we find that hole intersubband transitions occur at energies close to 165 and 117 meV for WSe$_2$ flakes with $N$ = 4 and $N$ = 5, respectively. Therefore, these spatial absorption maps are a first indication that we observe intersubband transitions. ii) The absorption in the monolayer region at $E_{ph}$ = 117 meV corresponds to Drude absorption: Since the charge carriers induced by the applied $V_{BG}$ are confined within a single atomic layer, the absorption due to free charge carriers is much higher than in the few-layer regions, where the same number of charge carriers is distributed over several layers (see SI chapter V). Thus, the relative contribution of Drude absorption is larger in the monolayer region.

To corroborate our interpretation, we spectrally resolve the intersubband resonances by changing $E_{ph}$ in small steps. We average the recorded optical signals $s_3$ over the areas of the flake with a constant thickness. Due to the interferometric detection, $s_3$ is a complex signal and allows us to recover both the real and imaginary part of the relative permittivity of WSe$_2$, $\varepsilon_{WSe2}$, by using a thin-film inversion model (see SI chapter III) [37, 38]. We note that the flake thickness, the effect of the substrate, as well as different measurement parameters like tapping amplitude and demodulation order are accounted for by the inversion model, but the qualitative resonance features are directly visible in the raw data. Figure 3a shows both $\Delta\varphi_3$ and the obtained $Im(\varepsilon_{WSe2})$ when the WSe$_2$ flake is p-doped, assuming $Im(\varepsilon_{WSe2}) = 0$ for charge-neutral WSe$_2$ in the infrared frequency range. First, we identify a general decrease of $Im(\varepsilon_{WSe2})$ with $E_{ph}$ for all layer thicknesses. We attribute this behavior to Drude absorption of holes that we account for (see SI chapter V) by a standard complex Drude term $\varepsilon_{\text{Drude}}(\omega) = -\frac{D}{\omega\varepsilon_0}\frac{1}{i\tau^{-1}+\omega}$, with the vacuum permittivity $\varepsilon_0$, the Drude weight $D$, scattering time $\tau$, and $E_{ph} = \hbar\omega$. In order to quantitatively address the intersubband absorption resonances, we subtract the Drude contribution from the obtained $Im(\varepsilon_{WSe2})$ (Figure 3b). We clearly distinguish an absorption peak for $N = 4$, as well as enhanced absorption for $N = 5$ at low photon energies. A Gaussian fit to the $N = 4$ data yields a center energy of 167.5 ± 1.5 meV and a full width half maximum $\Gamma = 33 \pm 4$ meV. The obtained $Re(\varepsilon_{WSe2})$ shows a resonance behavior at the same energy, supporting our conclusion that we indeed observe a resonant absorption. The experimentally observed resonance energy (168 meV) is slightly higher than the theoretical prediction (145 meV and 154 meV for the two DFT calculations, respectively), which may be due to many-body effects [39] and corrections due to interactions with the electron or hole plasma [40].

We proceed with discussing the linewidth of the intersubband transitions. As 2D materials have atomically smooth surfaces, we expect that the linewidth of van der Waals quantum wells is not significantly broadened by interface roughness scattering, which is the dominant broadening mechanism in III-V semiconductor quantum wells [41, 42]. However, the linewidth can be broadened both by phonons or disorder. We calculate the optical-phonon-limited linewidth at low temperatures to be 0.66 meV and therefore much smaller than in epitaxial quantum wells [41]. At room temperature, the calculated linewidth $\Gamma \approx 5.9$ meV is much larger, which originates from the differences in in-plane effective masses in the different subbands. The holes in the second subband are heavier than in the first subband, spreading the absorption for high temperatures or strongly doped films to $\Gamma \approx \left[1 - \frac{m_1}{m_2}\right] \max\{\varepsilon_F, k_B T \log 2\}$, where $m_1$ and $m_2$ are the in-plane masses in the first and second subband respectively, $m_j^{-1} = m_{xy}^{-1}\left(1 + \frac{\pi^2 j^2 \zeta}{(N+2\nu)^2 d^2}\right)$, where the difference is determined by the non-parabolicity in the dispersion of holes in the bulk crystal, $\varepsilon_F$ is the Fermi energy, $k_B$ Boltzmann's constant and $T$ the temperature (see SI chapter II for more details). The measured linewidth of $\Gamma = 33$ meV is larger than these theoretical estimates for the intrinsic linewidth. An additional broadening mechanism in our experiment stems from the sharpness of the AFM tip, providing momentum for non-vertical transitions. In order to quantify this, we examine in more detail the hot spot around the AFM tip apex that excites intersubband transitions. The in-plane momenta contributing to the near-field coupling are given by a broad Bell shaped function (see SI chapter II) peaking at a k-vector that is inversely related to the tip curvature [43]. Due to this broad momentum distribution, a wide range of non-vertical transitions with finite momentum transfer and a distribution of transition energies contribute to the signal. This leads to a broadening and blue shift of the absorption line shape. The sharpest features around the AFM tip apex determine the in-plane momentum distribution and are typically around 12 nm for the used AFM tips. Calculations for this tip radius yield $\Gamma = 24$ meV (Figure 3c), indicating that tip-induced broadening plays an important role for the experimentally observed linewidth. However, we cannot exclude additional contributions to the linewidth broadening due to disorder.

Another key signature of intersubband transitions is the dependence of the sheet absorption $\alpha$ on the carrier density $n_{2D}$. $\alpha$ is typically obtained in far-field absorption measurements [9, 33]. For an ideal, infinitely deep quantum well, assuming parallel subbands and $T = 0$ K, calculations yield $\alpha = n_{2D} \frac{e^2 \hbar f_{12}}{c \varepsilon m_z \Gamma}$ [11], with the electron charge $e$, speed of light $c$, oscillator strength $f_{12} = 0.96$ for the transition from the first to the second subband, and $\varepsilon$ the static out-of-plane permittivity. To experimentally study this dependence, we vary $n_{2D}$ by changing the modulation amplitude of the backgate voltage, $V_{BG,max}$, at $E_{ph}$ = 165 meV, which corresponds to the center energy of the $N$ = 4 intersubband resonance. The measured (Drude-corrected) $Im(\varepsilon_{WSe2})$ as a function of $n_{2D}$, shown in Figure 4a, displays a monotonic increase of the absorption with $n_{2D}$. This is consistent with the idealized square well model, although we note that the exact function of the increase might be affected by possible contributions from many-body effects and non-parabolic bands. Experimentally, we find for the highest hole density ($n_{2D}$ = 9x10$^{11}$ cm$^{-2}$, extracted from capacitance measurements on a separate device) a maximum absorption of $\alpha_{exp}$ = 0.017 ± 0.002 %, which we calculated from the measured complex value of $\varepsilon_{WSe2}$ (see SI chapter VII). This is in good agreement with the idealized model yielding $\alpha$ = 0.026% for a WSe$_2$ crystal with $N$ = 4 taking into account the measured linewidth $\Gamma$ = 33 meV. We note that the absorption will be significantly enhanced with narrower linewidths, reaching values as high as α = 1.3% for the phonon-limited linewidth $\Gamma$ = 0.66 meV.

We further demonstrate the versatility of 2D materials by ambipolar electrical control of intersubband transitions in the valence and conduction bands. We tune the carrier density from the p-doped to the n-doped regime and observe intersubband transitions in both cases. Interestingly, at $E_{ph}$ = 117 meV, these transitions occur for different layer thicknesses. As we have seen before, we observe hole intersubband transitions in the $N$ = 5 area of the flake (Figure 4b). However, when the flake is n-doped, we observe enhanced absorption due to electron intersubband transitions in the $N$ = 6 region of the flake (Figure 4c). This observation that the electron transitions for a given $E_{ph}$ occur in a thicker part of the flake than the hole transitions is in excellent agreement with our theoretical ab-initio calculations (see Figures 1f-g). We note that in order to reach n- and p-doping without leakage through the underlying oxide we use the AFM tip to induce additional doping to the flake (see SI chapter VI).

In conclusion, we reported the first experimental observation of intersubband transitions in few-layer WSe$_2$ crystals with spatially and spectrally resolved measurements supported by detailed ab-initio calculations. The obtained spatial resolution of 20 nm cannot be obtained by classical far-field measurements. Intersubband transitions are not unique to WSe$_2$; they can be found in any gapped 2D material (see SI chapter VIII for similar measurements on MoS$_2$). This study represents a first step into a vast, unexplored field. In a future study, the role of disorder and many-body interactions needs to be investigated more thoroughly in order to provide conclusions on the intrinsic linewidth. Recently, wafer-scale multilayer TMD flakes have become commercially available and progress is being made to reduce the defect density in these large area films [44]. This provides the opportunity to also study intersubband transitions in the far-field and is promising for future experiments and applications. Our study offers a first glimpse of the physics and technology potentially enabled by van der Waals quantum wells, such as infrared detectors, sources, and lasers with the potential for compact integration with Si CMOS. Particularly appealing is the possibility of combining various 2D material QWs [14, 45] to any of the wide variety of different 2D crystals – including semimetals, dielectrics, topological insulators [46, 47], superconductors [48], and ferromagnets [49, 50] – with any relative lattice alignment angle between different layers, giving rise to an unprecedented freedom in designing novel optoelectronic devices. This paves the way to new phenomena that are not accessible in any other material class.

**Data and code availability statement:**

The raw data of this study as well as all the code used to process this data are available from the corresponding author upon request.

**References:**

[1] Kroemer, H. (2001). Nobel Lecture: Quasi-electric Fields and Band Offset: Teaching Electrons New Tricks. *Reviews of Modern Physics*, *73*(3), 783–793.

[2] Alferov, Z. I. (2001). Nobel lecture: The double heterostructure concept and its applications in physics, electronics, and technology. *Reviews of Modern Physics*, *73*(3), 767–782.

[3] Hayashi, I., Panish, M. B., Foy, P. W., & Sumski, S. (1970). Junction lasers which operate continuously at room temperature. *Applied Physics Letters*, *17*(3), 109–111.

[4] Nakamura, S., Mukai, T., & Senoh, M. (1994). Candela-class high-brightness InGaN/AlGaN double-heterostructure blue-light-emitting diodes. *Applied Physics Letters*, *64*(13), 1687–1689.

[5] Bhattacharya, P., & Mi, Z. (2007). Quantum-dot optoelectronic devices. *Proceedings of the IEEE*, *95*(9), 1723–1740.

[6] Wang, J., Gudiksen, M. S., Duan, X., Cui, Y., & Lieber, C. M. (2001). Highly polarized photoluminescence and photodetection from single indium phosphide nanowires. *Science*, *293*(5534), 1455–1457.

[7] Levine, B. F. (1993). Quantumwell infrared photodetectors. *Journal of Applied Physics*, *74*(8), R1–R81.

[8] Faist, J., Capasso, F., Sivco, D., Sirtori, C., Hutchinson, A. L., & Cho, A. Y. (1994). Quantum cascade laser. *Science*, *264*(5158), 553–556.

[9] West, L. C., & Eglash, S. J. (1985). First observation of an extremely large-dipole infrared transition within the conduction band of a GaAs quantum well. *Applied Physics Letters*, *46*(12), 1156-1158.

[10] Helm, M. (2004). Intersubband semiconductor light sources: history, status, and future. In *Infrared and Millimeter Waves, Conference Digest of the 2004 Joint 29th International Conference on 2004 and 12th International Conference on Terahertz Electronics, 2004.*

[11] Liu, H. C., & Capasso, F. (2000). Intersubband transitions in quantum wells: Physics and device applications I. In *Semiconductors and Semimetals* (Vol. 62, p. 323).

[12] Warwick, C. A., Jan, W. Y., Ourmazd, A., & Harris, T. D. (1990). Does luminescence show semiconductor interfaces to be atomically smooth? *Applied Physics Letters*, *56*(26), 2666–2668.

[13] Novoselov, K. S., Jiang, D., Schedin, F., Booth, T. J., Khotkevich, V. V, Morozov, S. V, & Geim, a K. (2005). Two-dimensional atomic crystals. *Proceedings of the National Academy of Sciences of the United States of America*, *102*(30), 10451–10453.

[14] Novoselov, K. S., Mishchenko, A., Carvalho, A., & Castro Neto, A. H. (2016). 2D materials and van der Waals heterostructures. *Science*, *353*(6298), 461.

[15] Mak, K. F., & Shan, J. (2016). Photonics and optoelectronics of 2D semiconductor transition metal dichalcogenides. *Nature Photonics*, *10*(4), 216–226.

[16] Wang, Q. H., Kalantar-Zadeh, K., Kis, A., Coleman, J. N., & Strano, M. S. (2012). Electronics and optoelectronics of two-dimensional transition metal dichalcogenides. *Nature Nanotechnology*, *7*(11), 699–712.

[17] Koppens, F. H. L., Mueller, T., Avouris, P., Ferrari, A. C., Vitiello, M. S., & Polini, M. (2014). Photodetectors based on graphene, other two-dimensional materials and hybrid systems. *Nature Nanotechnology*, *9*(10), 780–793.

[18] Cui, X., Lee, G.-H., Kim, Y. D., Arefe, G., Huang, P. Y., Lee, C.-H., … Hone, J. (2015). Multi-terminal transport measurements of $MoS_2$ using a van der Waals heterostructure device platform. *Nature Nanotechnology*, *10*(6), 534–540.

[19] Xu, S., Shen, J., Long, G., Wu, Z., Bao, Z. Q., Liu, C. C., … Wang, N. (2017). Odd-Integer Quantum Hall States and Giant Spin Susceptibility in p-Type Few-Layer $WSe_2$. *Physical Review Letters*, *118*, 067702.

[20] Bandurin, D. A., Tyurnina, A. V., Yu, G. L., Mishchenko, A., Zólyomi, V., Morozov, S. V., … Cao, Y. (2017). High electron mobility, quantum Hall effect and anomalous optical response in atomically thin InSe. *Nature Nanotechnology*, *12*(3), 223–227.

[21] Schaibley, J. R., Yu, H., Clark, G., Rivera, P., Ross, J. S., Seyler, K. L., … Xu, X. (2016). Valleytronics in 2D materials. *Nature Reviews Materials*, *1*(11), 1–15.

[22] Mak, K. F., McGill, K. L., Park, J., & McEuen, P. L. (2014). The valley Hall effect in $MoS_2$ transistors. *Science*, *344*(6191), 1489–92.

[23] Back, P., Zeytinoglu, S., Ijaz, A., Kroner, M., & Imamoğlu, A. (2018). Realization of an Electrically Tunable Narrow-Bandwidth Atomically Thin Mirror Using Monolayer $MoSe_2$. *Physical Review Letters*, *120*, 037401.

[24] Scuri, G., Zhou, Y., High, A. A., Wild, D. S., Shu, C., De Greve, K., … Park, H. (2018). Large Excitonic Reflectivity of Monolayer $MoSe_2$ Encapsulated in Hexagonal Boron Nitride. *Physical Review Letters*, *120*, 037402.

[25] Cadiz, F., Courtade, E., Robert, C., Wang, G., Shen, Y., Cai, H., … Urbaszek, B. (2017). Excitonic linewidth approaching the homogeneous limit in $MoS_2$-based van der Waals heterostructures. *Physical Review X*, *7*, 021026.

[26] Ajayi, O. A., Ardelean, J. V., Shepard, G. D., Wang, J., Antony, A., Taniguchi, T., … Hone, J. C. (2017). Approaching the intrinsic photoluminescence linewidth in transition metal dichalcogenide monolayers. *2D Materials*, *4*(3)

[27] Wang, L., Meric, I., Huang, P. Y., Gao, Q., Gao, Y., Tran, H., … Dean, C. R. (2013). One-dimensional electrical contact to a two-dimensional material. *Science*, *342*(6158), 614–7

[28] Britnell, L., Gorbachev, R. V., Jalil, R., Belle, B. D., Schedin, F., Mishchenko, A., … Ponomarenko, L. A. (2012). Field-effect tunneling transistor based on vertical graphene heterostructures. *Science*, *335*(6071), 947–950.

[29] Keilmann, F., & Hillenbrand, R. (2004). Near-field microscopy by elastic light scattering from a tip. *Phil. Trans. R. Soc. A*, *362*(1817), 787–805.

[30] Kormányos, A., Burkard, G., Gmitra, M., Fabian, J., Zólyomi, V., Drummond, N. D., & Fal'ko, V. (2014). kp theory for two-dimensional transition metal dichalcogenide semiconductors. *2D Materials*, *2*, 22001.

[31] Sahin, H., Tongay, S., Horzum, S., Fan, W., Zhou, J., Li, J., … Peeters, F. M. (2013). Anomalous Raman spectra and thickness-dependent electronic properties of $WSe_2$. *Physical Review B*, *87*(16), 165409.

[32] Huang, W., Luo, X., Gan, C. K., Quek, S. Y., & Liang, G. (2014). Theoretical study of thermoelectric properties of few-layer $MoS_2$ and $WSe_2$. *Physical Chemistry Chemical Physics*, *16*(22), 10866.

[33] Kane, M. J., Emeny, M. T., Apsley, N., Whitehouse, C. R., & Lee, D. (1988). Inter-subband absorption in GaAs/AlGaAs single quantum wells. *Semiconductor Science and Technology*, *3*, 722–725.

[34] Ocelic, N., Huber, A., & Hillenbrand, R. (2006). Pseudoheterodyne detection for background-free near-field spectroscopy. *Applied Physics Letters*, *89*, 101124.

[35] Huth, F., Govyadinov, A., Amarie, S., Nuansing, W., Keilmann, F., & Hillenbrand, R. (2012). Nano-FTIR absorption spectroscopy of molecular fingerprints at 20 nm spatial resolution. *Nano Letters*, *12*(8), 3973–3978.

[36] Taubner, T., Hillenbrand, R., & Keilmann, F. (2004). Nanoscale polymer recognition by spectral signature in scattering infrared near-field microscopy. *Applied Physics Letters*, *85*(21), 5064–5066.

[37] Govyadinov, A. A., Amenabar, I., Huth, F., Carney, P. S., & Hillenbrand, R. (2013). Quantitative measurement of local infrared absorption and dielectric function with tip-enhanced near-field microscopy. *The Journal of Physical Chemistry Letters*, *4*(9), 1526–1531.

[38] Govyadinov, A. A., Mastel, S., Golmar, F., Chuvilin, A., Carney, P. S., & Hillenbrand, R. (2014). Recovery of permittivity and depth from near-field data as a step toward infrared nanotomography. *ACS Nano*, *8*(7), 6911–6921.

[39] Manasreh, M. O., Szmulowicz, F., Vaughan, T., Evans, K. R., Stutz, C. E., & Fischer, D. W. (1991). Origin of the Blueshift in the Intersubband Infrared Absorption in $GaAs/Al_{0.3}Ga_{0.7}As$ Multiple Quantum Well. *Physical Review B*, *43*(12), 9996–9999.

[40] Allen, S. J., Tsui, D. C., & B, V. (1993). On the absorption of infrared radiation by electrons in semiconductor inversion layers. *Solid State Communications*, *88*, 425-428.

[41] Unuma, T., Yoshita, M., Noda, T., Sakaki, H., & Akiyama, H. (2003). Intersubband absorption linewidth in GaAs quantum wells due to scattering by interface roughness, phonons, alloy disorder, and impurities. *Journal of Applied Physics*, *93*(3), 1586–1597.

[42] Tsujino, S., Borak, A., Müller, E., Scheinert, M., Falub, C. V., Sigg, H., … Faist, J. (2005). Interface-roughness-induced broadening of intersubband electroluminescence in p-SiGe and n-GaInAs/AlInAs quantum-cascade structures. *Applied Physics Letters*, *86*, 062113.

[43] Fei, Z., Andreev, G. O., Bao, W., Zhang, L. M., S. McLeod, A., Wang, C., ... Basov, D. N. (2011). Infrared nanoscopy of Dirac plasmons at the graphene-$SiO_2$ interface. *Nano Letters*, *11*, 4701-4705

[44] Edelberg, D., Rhodes, D., Kerelsky, A., Kim, B., Wang, J., Zangiabadi, A., … Balicas, L. (2018). Hundredfold Enhancement of Light Emission via Defect Control in Monolayer Transition-Metal Dichalcogenides. *arXiv:1805.00127*

[45] Geim, A. K., & Grigorieva, I. V. (2013). Van der Waals heterostructures. *Nature*, *499*(7459), 419–25.

[46] Fei, Z., Palomaki, T., Wu, S., Zhao, W., Cai, X., Sun, B., … Cobden, D. H. (2017). Edge conduction in monolayer $WTe_2$. *Nature Physics*, *13*(7), 677–682.

[47] Wu, S., Fatemi, V., Gibson, Q. D., Watanabe, K., Taniguchi, T., Cava, R. J., & Jarillo-Herrero, P. (2018). Observation of the Quantum Spin Hall Effect up to 100 Kelvin in a Monolayer Crystal. *Science*, *359*(6371), 76–79.

[48] Xi, X., Wang, Z., Zhao, W., Park, J.-H., Law, K. T., Berger, H., … Mak, K. F. (2015). Ising pairing in superconducting $NbSe_2$ atomic layers. *Nature Physics*, *12*(2), 139–143.

[49] Huang, B., Clark, G., Navarro-Moratalla, E., Klein, D. R., Cheng, R., Seyler, K. L., … Xu, X. (2017). Layer-dependent ferromagnetism in a van der Waals crystal down to the monolayer limit. *Nature*, *546*(7657), 270–273.

[50] Gong, C., Li, L., Li, Z., Ji, H., Stern, A., Xia, Y., … Zhang, X. (2017). Discovery of intrinsic ferromagnetism in two-dimensional van der Waals crystals. *Nature*, *546*(7657), 265–269.

**Acknowledgements:**

We greatly acknowledge discussions with Prof. Simon Wall about the experimental measurement technique. We also like to thank Dr. Alexander Govyadinov on discussions about the thin-film inversion model. P.S. acknowledges financial support by a scholarship from the 'la Caixa' Banking Foundation. F.V. acknowledges financial support from Marie-Curie International Fellowship COFUND and ICFOnest program. M.M. thanks the Natural Sciences and Engineering Research Council of Canada (PGSD3-426325-2012). K.-J.T. acknowledges support from a Mineco Young Investigator Grant (FIS2014-59639-JIN). F.K. acknowledges financial support from the Government of Catalonia trough the SGR grant (2014-SGR-1535), and from the Spanish Ministry of Economy and Competitiveness, through the "Severo Ochoa" Programme for Centres of Excellence in R&D (SEV-2015-0522), support by Fundacio Cellex Barcelona, CERCA Programme / Generalitat de Catalunya and the Mineco grants Ramón y Cajal (RYC-2012-12281) and Plan Nacional (FIS2013-47161-P and FIS2014-59639-JIN). Furthermore, the research leading to these results has received funding from the European Union Seventh Framework Programme under grant agreement no.696656 Graphene Flagship, ERC Starting grant (307806, CarbonLight) and ERC Synergy Grant Hetero2D.

## Figure captions:

**Figure 1 | Van der Waals quantum wells – concept and theory. a**, Schematic illustration of charge carriers (in blue) confined within a TMD flake consisting of different thicknesses. **b**, Upon light excitation with a photon energy $E_{ph}$ and an out-of-plane polarization, charge carriers can be excited from the ground state to the first excited state (in pink), if the intersubband transition energy for a given thickness is resonant with $E_{ph}$. **c**, Ab-initio DFT band structure calculations for a $N = 5$ 2H-WSe$_2$ crystal. Shown are the five highest valence bands and the five lowest conduction bands, displayed along the Γ-K axis of the hexagonal Brillouin zone as indicated in the inset. SO coupling leads to an additional splitting of the conduction bands. The lowest three conduction bands can be assigned to spin down (up) and are marked by solid (dashed) lines. **d**, **e**, Calculated out-of-plane wave functions for a WSe$_2$ crystal with $N = 5$, at the points in the Brillouin zone as indicated in **c**. Panel **d** shows the electron wave functions of the three lowest conduction bands at the Λ point, and panel **e** shows the hole wave functions of the three highest valence bands at the Γ point. Dashed lines represent the envelope functions assuming a perfect, infinitely deep square well potential. **f**, **g**, Calculated transition energies $E_{sub}$ from the first to the second subband as indicated by the purple arrows in **d** and **e** as a function of $N$ for electrons in the conduction band (**f**) and holes in the valence band (**g**). Blue (red) squares correspond to transitions between the spin down (up) polarized subbands, whereas purple circles correspond to transitions between spin-degenerate subbands. The purple dashed line is a fit to the spin-degenerate hole transitions using the modified infinite square well model (eq. 1) yielding $v = 0.09$.

**Figure 2 | Measurement setup and spatial absorption maps of a terraced WSe$_2$ flake. a**, Real colour microscope image of the WSe$_2$ flake and schematic illustration of the s-SNOM (NeaSNOM from Neaspec GmbH) measurement setup. A MIR laser beam is divided into two paths by a beam splitter (BS) for interferometric detection: One path is reflected by a mirror oscillating with $f_{osc} \approx 300$ Hz, while the other beam is focused on the apex of a metalized AFM tip oscillating with $f_{tip} \approx 250$ kHz, where it interacts with the sample in the near-field. The back scattered light carries information about the sample permittivity at the illumination photon energy $E_{ph}$. Both beams are recombined and detected by a cooled HgCdTe (MCT) detector. In order to suppress the large background of light that is reflected from the AFM tip without interacting with the sample, the signal is evaluated at sidebands (arising due to the mirror oscillation with $f_{osc}$) of higher harmonics $n$ of $f_{tip}$ by a built-in lock-in amplifier (LIA) [34]. Due to the interferometric detection, the recorded signal $s_n$ is complex and for our specific system of a thin WSe$_2$ flake on top of a thick substrate, $|s_n|$ increases monotonically with the number of layers $N$ of the WSe$_2$ flake (see SI). $s_n$ is further modulated by applying a square wave backgate voltage $V_{BG}$ with frequency $f_{BG}$ ($f_{BG} << f_{osc} << f_{tip}$) to the Si substrate thus modulating the doping of the WSe$_2$ flake. The change in phase of $s_n$ with doping, $\Delta\varphi_n$, is proportional to the change in the imaginary part of the permittivity of WSe$_2$, $\Delta Im(\varepsilon_{WSe2})$. **b**, Spatial map of $|s_3|$ in arbitrary units obtained by scanning the AFM tip over the area of the flake marked by the dashed lines in **a**. Different layer numbers can be identified by their optical signal and are indicated in the map. **c**, Time trace of the modulated backgate voltage $|V_{BG}|$ (black solid line, left axis) and corresponding phase of the third harmonic detector signal $\varphi_3$ (red dots, right axis). Data obtained for $E_{ph} = 117$ meV in the resonant $N = 5$ layer of a WSe$_2$ flake with with $f_{BG} = 1.9$ Hz. Dashed red lines indicate the average of $\varphi_3$ for when the flake is p-doped or charge neutral. The difference between these averages corresponds to $\Delta\varphi_3$. **d**, $\Delta\varphi_3$ on the same flake as in **c**, which contains terraces of different $N$. The data was obtained at $E_{ph} = 117$ meV by averaging 10 identical line scans

to increase the signal/noise ratio. **e**, Spatial map of $\Delta\varphi_3$ obtained during the same scan as shown in **b** with $E_{ph}$ = 117 meV. Higher values of $\Delta\varphi_3$ correspond to higher absorption. **f**, Same as **e** but for $E_{ph}$ = 165 meV.

**Figure 3 | Intersubband absorption spectra in few-layer WSe$_2$. a**, Experimentally obtained $Im(\varepsilon_{WSe2})$ (filled dots) for different excitation photon energies $E_{ph}$ and layer numbers $N$. $Im(\varepsilon_{WSe2})$ was calculated from the measured optical signals using a thin-film inversion model [37, 38]. The measurement was performed for a hole concentration $n_{2D} \approx$ 9x10$^{11}$ cm$^{-2}$. The yellow dashed line is a Drude fit to the $N$ = 6 data. To demonstrate that the intersubband resonance for $N$ = 4 is also visible in the raw data, we show $\Delta\varphi_3$ for $N$ = 4 (open circles, right axis) **b**, Same data as in **a** with the Drude contribution subtracted. The blue dashed line shows a Gaussian fit to the $N$ = 4 intersubband resonance, yielding a center energy of 167.5 ± 1.5 meV and linewidth $\Gamma$ = 33 ± 4 meV. The orange dashed line is a Gaussian guide to the eye for the $N$ = 5 intersubband resonance, assuming the same tip-induced $\Gamma$. Inset: $Re(\varepsilon_{WSe2})$ for the $N$ = 4 area of the flake. The dashed line is a guide to the eye for a Lorentzian resonance. **c**, Absorption line shapes obtained from *ab-initio* calculations. Calculations were done for low carrier concentrations (Boltzmann distribution) at the $\Gamma$ point of a p-doped 2H-WSe$_2$ crystal with $N$ = 4 (see SI). The solid blue line shows the expected line shape for transitions excited by the near-field around an AFM tip with a radius of 12 nm at room temperature. The line shape is broadened and blue-shifted compared to far-field excitations with negligible in-plane momentum, shown as dashed lines for $T$ = 300 K (black) and $T \sim$ 0 K (purple). Insets illustrate the possible transitions from the first to the second subband. The two subbands exhibit slightly different effective masses.

**Figure 4 | Doping dependence of electron and hole intersubband absorption. a,** Drude corrected $Im(\varepsilon_{WSe2})$ for different hole densities $n_{2D}$. The measurement was done at $E_{ph}$ = 165 meV, which corresponds to the centre of the $N$ = 4 intersubband resonance. The carrier concentration was controlled by varying the maximum voltage $V_{BG,max}$ of the square wave function applied to the backgate. Insets show possible transitions for low and high carrier densities. **b, c,** Observation of hole (**b**, for $N$ = 5 in orange) and electron (**c**, for $N$ = 6 in yellow) intersubband transitions within the same WSe$_2$ flake at $E_{ph}$ = 117 meV. Shown is the Drude corrected $Im(\varepsilon_{WSe2})$. The DC backgate voltage $V_{BG}$ was increased stepwise from -30 to +20 V. The measurement was performed on a WSe$_2$ flake on top of a SiO$_2$ substrate. To reach higher doping levels, an additional DC voltage $V_{tip}$ was applied between the WSe$_2$ flake and the AFM tip, locally inducing holes (**b**, for $V_{tip}$ = +3 V) or electrons (**c**, for $V_{tip}$ = -3 V) due to interactions with the metallized, grounded AFM tip.

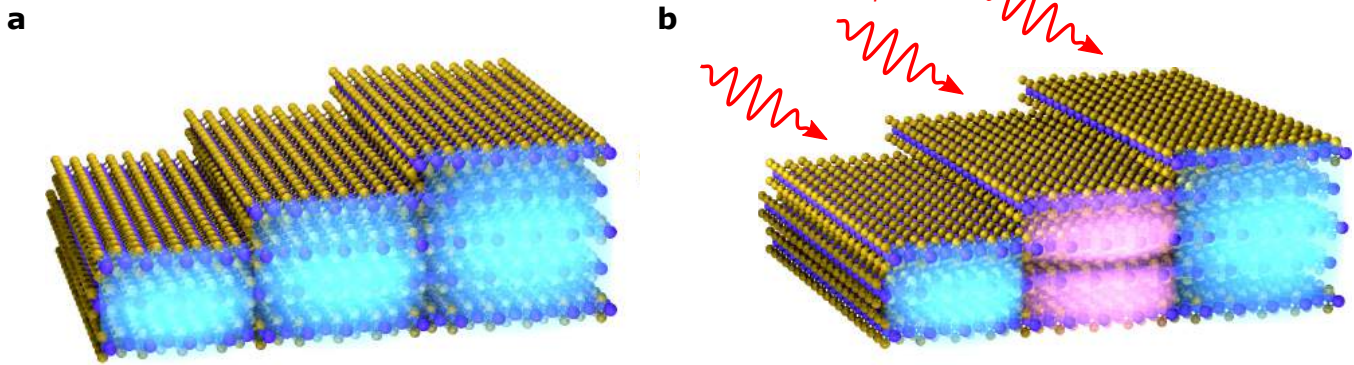
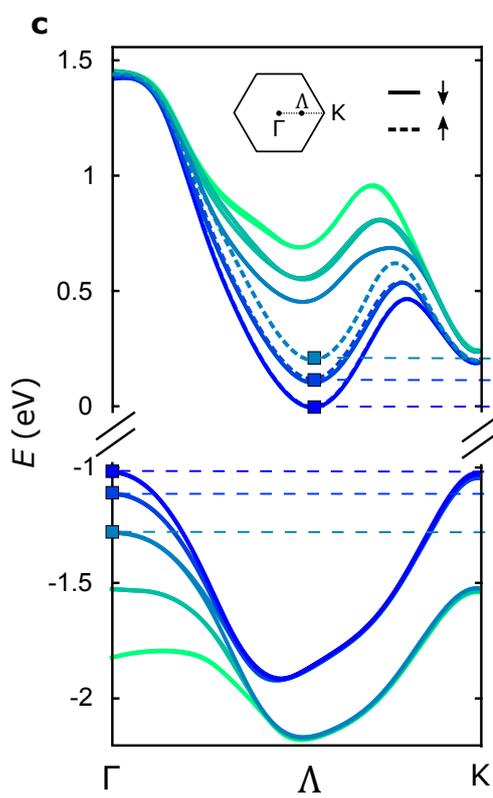
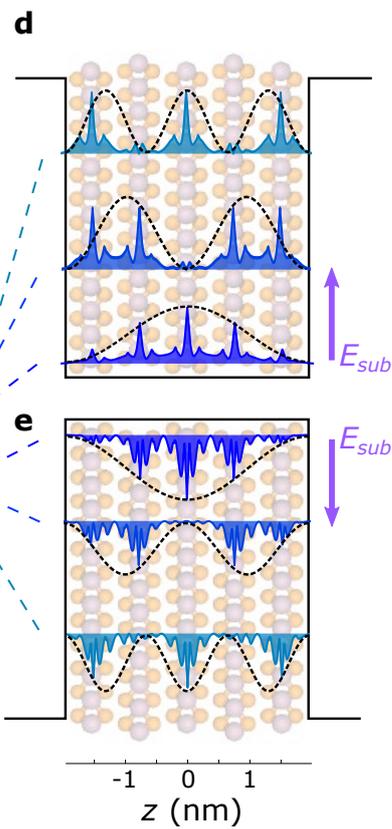
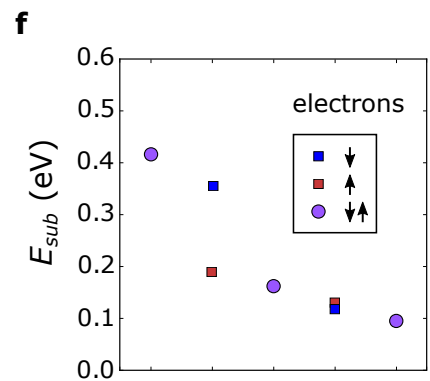
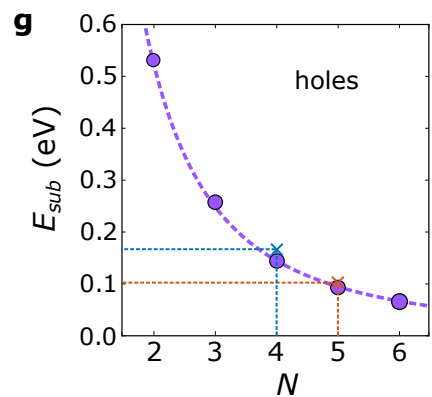

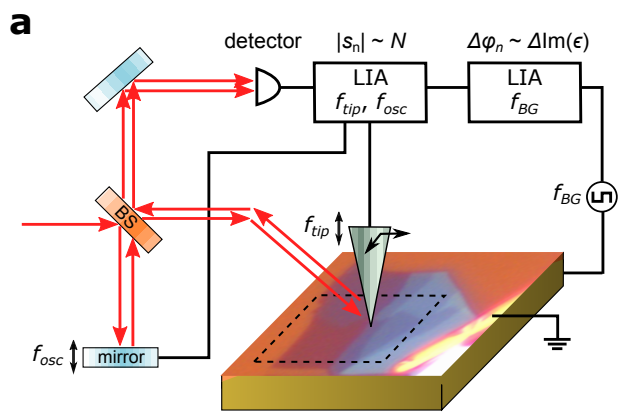
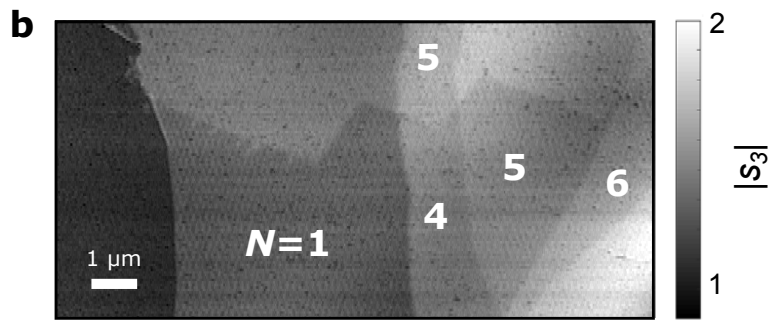
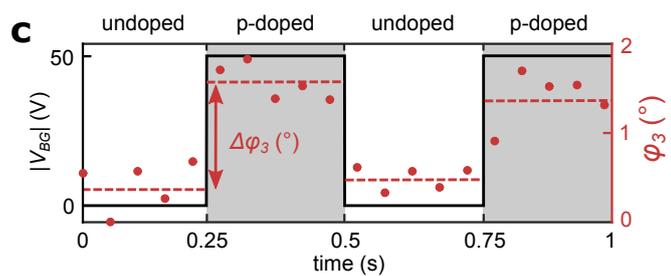
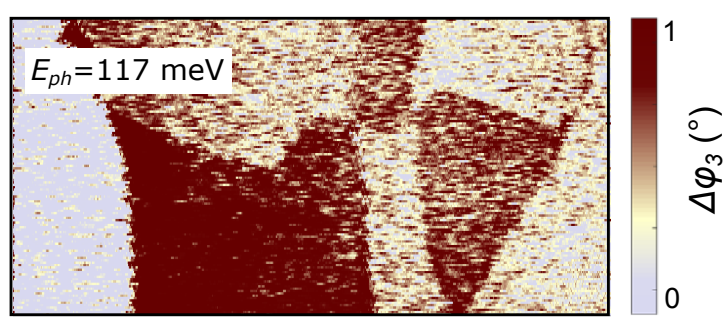
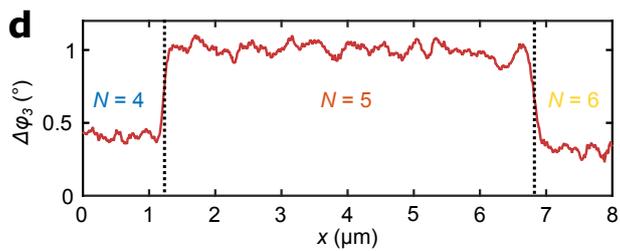
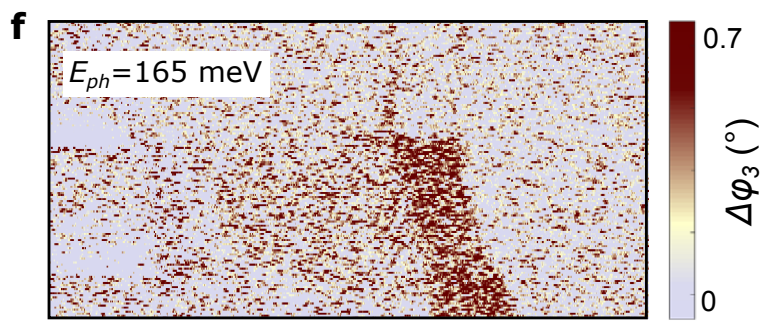

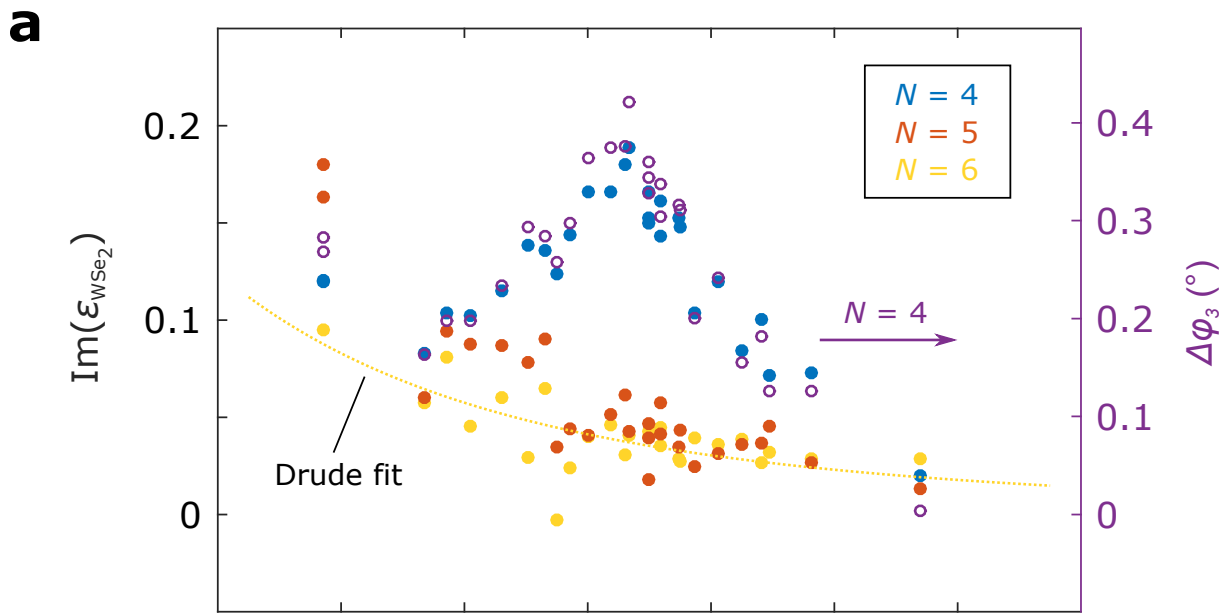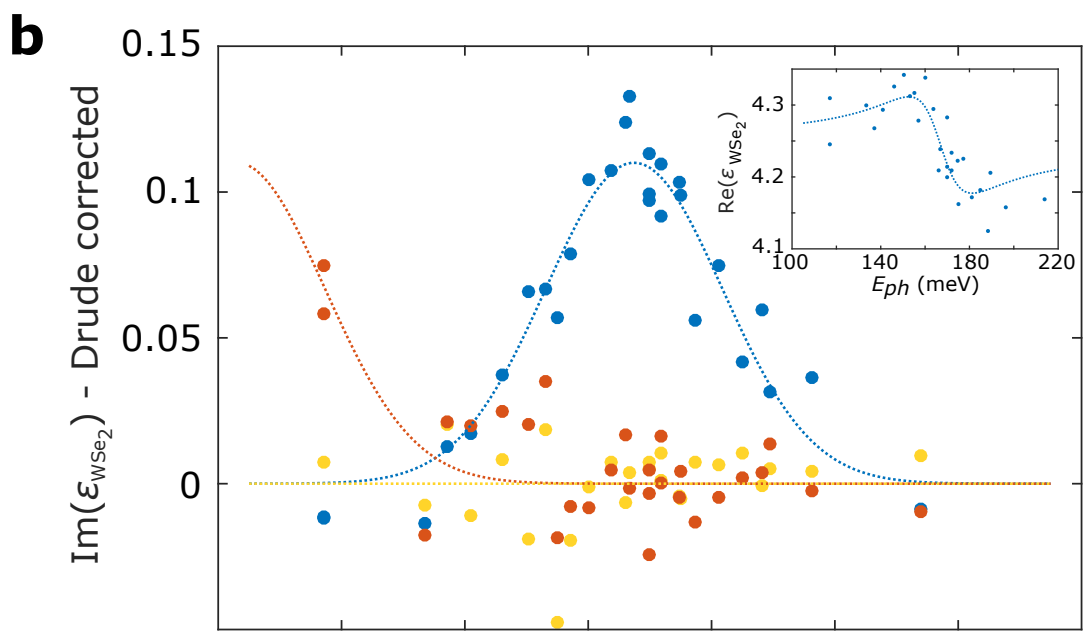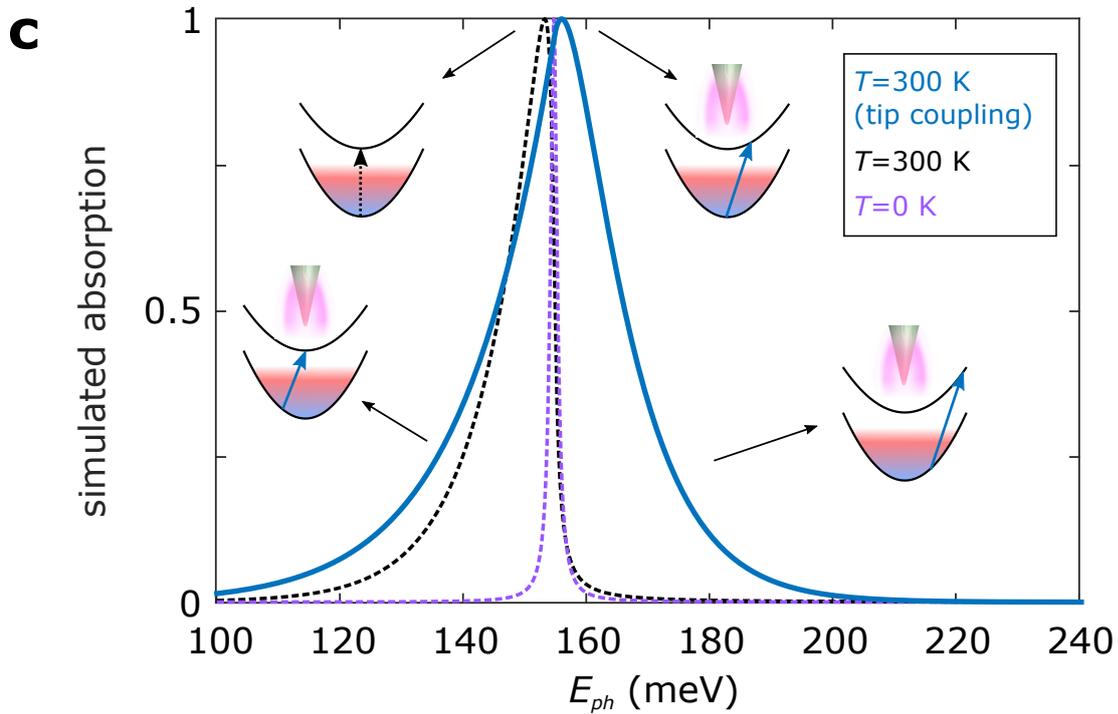

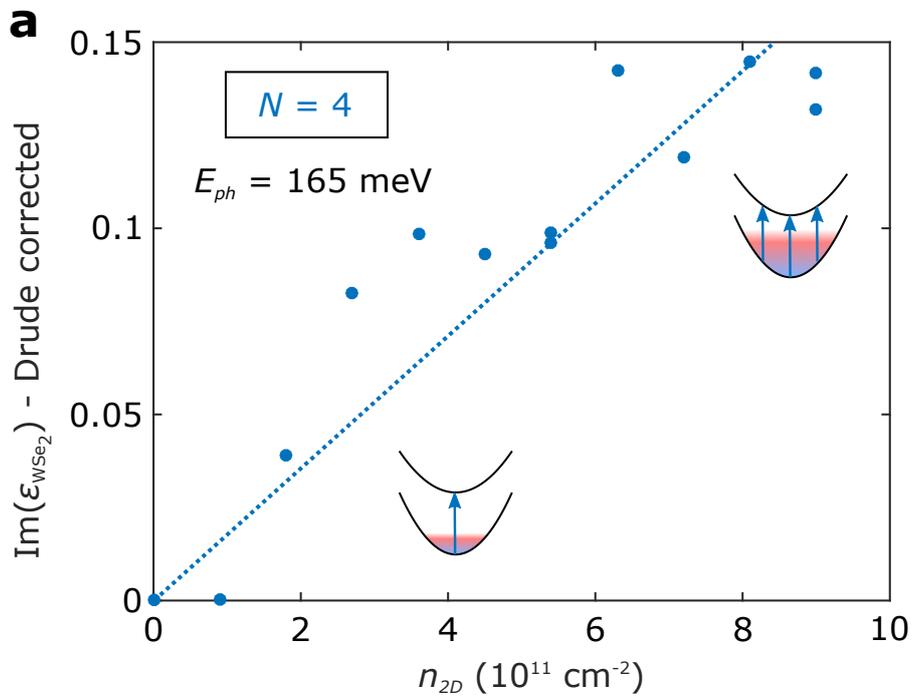

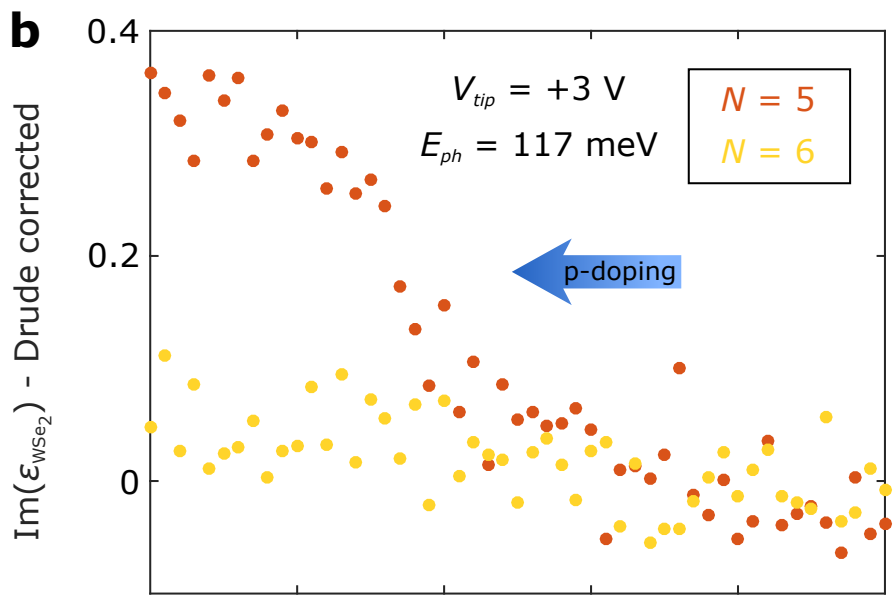

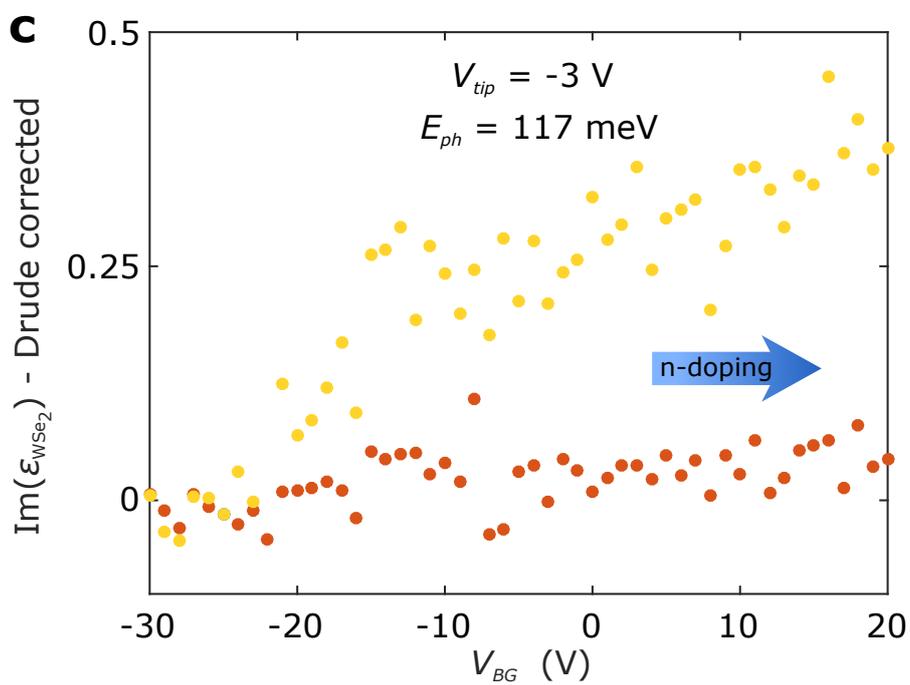